\documentclass[10pt,twocolumn,letterpaper]{article}

\usepackage[pagenumbers]{cvpr} 
\usepackage{tabu}
\usepackage{multirow}
\usepackage{subcaption}
\usepackage{adjustbox}
\usepackage[accsupp]{axessibility}
\usepackage{amsmath}
\usepackage{epigraph}

\usepackage{tikz}
\usepackage{pgfplots}
\pgfplotsset{compat=default}
\usetikzlibrary{pgfplots.groupplots}

\usepackage{adjustbox}
\usepackage{tabu}
\usepackage{multirow}
\usepackage{subcaption}

\newcommand{\TODO}[1]{\textbf{\color{red}[TODO: #1]}}
\renewcommand{\TODO}[1]{}

\definecolor{cvprblue}{rgb}{0.21,0.49,0.74}
\usepackage[pagebackref,breaklinks,colorlinks,citecolor=cvprblue]{hyperref}

\def\paperID{150} 
\def\confName{CVPR}
\def\confYear{2025}

\title{Even the ``Devil'' Has Rights!}

\author{Mennatullah Siam\\
Ontario Tech University, ON, Canada}

\begin{document}

\makeatletter
\DeclareRobustCommand\onedot{\futurelet\@let@token\@onedot}
\def\@onedot{\ifx\@let@token.\else.\null\fi\xspace}
\def\eg{\emph{e.g}\onedot} \def\Eg{\emph{E.g}\onedot}
\def\ie{\emph{i.e}\onedot} \def\Ie{\emph{I.e}\onedot}
\def\cf{\emph{cf}\onedot} \def\Cf{\emph{C.f}\onedot}
\def\etc{\emph{etc}\onedot} \def\vs{\emph{vs}\onedot}
\def\wrt{w.r.t\onedot} \def\dof{d.o.f\onedot}
\def\etal{\emph{et al}\onedot}

\def\latentregion{latent region}
\def\latentregionrep{latent token}

\makeatother
\maketitle

There have been works discussing the adoption of a human rights framework for responsible AI, emphasizing various rights such as the right to contribute to scientific advancements. Yet, to the best of our knowledge, this is the first attempt to take this framework with special focus on computer vision and documenting human rights violations in its community. This work summarizes such incidents accompanied with evidence from the lens of a female African Muslim Hijabi researcher. While previous works resorted to qualitative surveys that gather opinions from various researchers in the field, this work argues that a single documented violation is sufficient to warrant attention regardless of the stature of this researcher. Incidents documented in this work include silence on Genocides that are occurring while promoting the governments contributing to it, a broken reviewing system and corruption in the faculty support systems. 
This work discusses that demonizing individuals for discrimination based on gender, ethnicity, creed or reprisal has been a successful tool for exclusion with documented evidence from a single case. We argue that human rights are guaranteed for every single individual even the ones that might be labelled as devils in the community for whichever reasons to dismantle such a tool from its roots.


\section{Introduction}
\epigraph{\textbf{TL;DR. We used to be good people until we wanted to become successful.}}

\begin{figure}[t!]
    \includegraphics[width=0.48\textwidth]{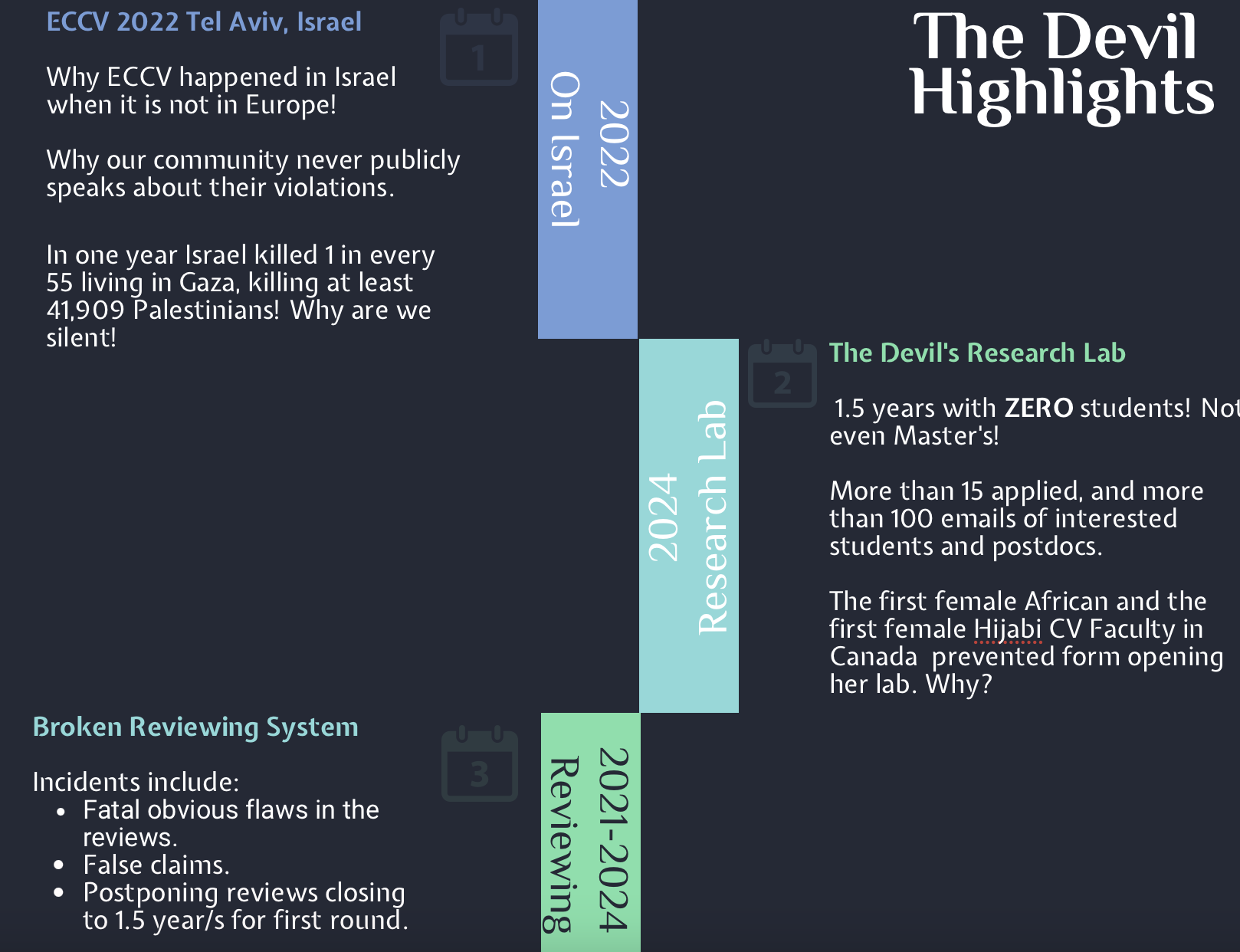}
    \caption{An overview of the ``devil'' highlights. The devil, is a female African Muslim Hijabi researcher that is actively publishing in the field. While that could be a shocking or funny label for some, but the devil label has been stamped on many researchers over the years towards expelling them from the field. This work is an effort to defend the right of every individual to contribute to research unless there is compelling evidence of unethical behavior. Even if a researcher is ``bad'', this researcher has the right to be rejected fairly to help them improve. Even if there is a researcher that is labelled to harm others, they should still be granted transparency; to be fully informed of the allegations and given a chance to defend themselves. Demonizing members of marginalized communities has, in our belief, been a highly effective tool for discrimination in our community. This work seeks to defend the idea that even those labeled as ``devils'' have rights; challenging and dismantling a tool of exclusion from its roots.}
    \label{fig:overview}
\end{figure}

Computer vision research has few works that discusses the research community itself beyond the traditional focus on building and developing algorithms and datasets~\cite{su2021affective,omotayo2023towards,omotayo2024state}. Over the years, the field has expanded, from classical computer vision to deep learning breakthroughs and now to foundation models~\cite{bommasani2021opportunities}. With this growth, there is an expected emotional impact on researchers~\cite{su2021affective}. However, previous work has primarily documented the perspectives of White and Asian researchers, who represent the majority in the field in terms of ethnicity. Researchers from underrepresented groups often fear retaliation and reprisals, making them reluctant to share experiences that expose corruption within the community. This work seeks to move beyond those limitations by documenting human rights violations on a female African Muslim Hijabi researcher, which eventually led her to practically leave her job, despite her contributions to top venues in her field (including ten publications in ICCV, CVPR, TPAMI, ICRA, IROS, and IJCAI). While some may argue that the experience of one researcher should not warrant the community’s attention. Within a human rights framework, the existence of even one documented violation with supporting evidence should be sufficient to drive the community toward a solution regardless of the stature of this researcher. 


First, let's re-iterate on the importance of a human rights framework that has been previously revisited towards a responsible AI~\cite{prabhakaran2022human}. It encompasses major rights like the right to life, liberty and security but also the right to ``share in scientific advancement and its benefits'' (UDHR, Art. 27)~\cite{prabhakaran2022human}. Under the assumption and quoting~\cite{prabhakaran2022human} ``to center the conversation around who is harmed, what harms they face, and how those harms may be mitigated.'', this work is documenting the experience of a researcher who was sufficiently harmed to practically leave her job. It documents various forms of harm as an author, as a reviewer, as a postdoctoral fellow and as a faculty. That researcher has been silenced over the years even when evidence was conveyed to researchers with high stature in the field, yet there was no support in solving these issues. It brings the question of why certain researchers and their corresponding sub-groups within diversity are treated differently and silenced. 

This work puts even more emphasis on such ``selective diversity'', where not all racialized individuals are treated the same. It is a well established norm that human rights are guaranteed to all individuals, even prisoners have rights to be respected. So let's reiterate that each researcher in the community has the right to contribute to computer vision research governed by the normative standards of the field, such as the reviewing guidelines and the ethical guidelines to conducting research. 
\textit{The only legitimate barrier to participating in any research community should be the presence of strong and compelling evidence of unethical behavior, whether in conducting research or in interactions with fellow researchers. Yet it has to be provided with transparency and allowing that researcher with the right to defend themselves.} While reviewing and rejecting publications of low quality might seem as a barrier, except if implemented properly it is indeed an enabler to help researchers improve their work. However, if not implemented properly it turns into an exclusion tool. This work discusses that even the devil has rights, whether being demonized for the right or wrong reasons, to avoid the use of demonizing racialized individuals to exclude them in any community. Otherwise, we can sell that our community members are pursuing success with whichever means even if it entails unfairly excluding researchers, for the sake of competition. The incidents documented in this work if not addressed will result in two major outcomes: (i) normalizing such behavior in the field, (ii) killing the right to freedom of expression and turning our community into a dictatorship that does not welcome critics. All documented incidents are provided anonymized to protect the privacy of individuals involved in this story beyond the discussed researcher.

The major issues that researcher witnessed over the years from her own lens include: (i) One of the major conferences, ECCV 2022, happening in a place with long history of human rights violations for people of her same religion and ethnicity, Sec.~\ref{sec:israel}. (ii) Being denied access to every opportunity to start her research lab and staying for a year and a half with zero students even when having two affiliations and grants, Sec.~\ref{sec:faculty}. (iii) Being denied from any position in universities that have computer vision labs actively publishing in the field within Canada, even when she holds dual citizenship as Canadian and when identified as a `Strong' researcher from their funding agency\footnote{\url{https://www.dropbox.com/scl/fi/99h8c40snkn8trx7sm6bj/DG_Results.pdf?rlkey=v16lrrnoghmls6md3lgyp0jt6&st=zaryt25b&dl=0}}. Even when other faculty that have lower publications quality, citations and identified as foreign workers were successfully appointed, Sec.~\ref{sec:faculty}. (iv) Being humiliated in various ways in front of undergraduate and graduate students, Sec.~\ref{sec:faculty}. (v) Being put under xxxxxxxxxxxxx three times in less than two years for no real reason. (vi) Papers getting rejected for obvious false or irrelevant claims and when requesting an appeal getting denied of that appeal, Sec.~\ref{sec:reviewing}. (vii) Spreading lies around that researcher without providing evidence, Sec.~\ref{sec:rumours}. Fig.~\ref{fig:overview} summarizes the major highlights of the researcher that is referred to throughout the work as the ``devil''.

\textit{It is important to note that while this researcher may be considered relatively average, being young and still in the process of improving her work, this does not justify such violations to go unnoticed. Human rights must be upheld for everyone, not just those with high status in the field.} This is especially crucial for young researchers trying to establish themselves. Despite being early in her career, this researcher has produced good contributions, relative to her diversity group, through high-quality publications. Some like to frame these incidents as ``normal'', claiming that ``we are all in the same boat'', except we are not. 

The following are incidents that this researcher has witnessed and had to endure while pursuing her studies and conducting research in the early stages of her career:
(i) She witnessed a massacre.
(ii) She was shot at with live ammunition by police and military during peaceful protests.
(iii) She participated in a peaceful sit-in defending her graduate university, where students were confronted by men who appeared to be thugs not wearing police uniforms.
(iv) She experienced multiple instances of police brutality in her home country and witnessed domestic violence against her mother by a male family member.
(v) She saw firsthand the ongoing human rights violations in Gaza as part of a humanitarian aid mission that included doctors and community volunteers like herself at Al-Shifaa hospital.
This is the context in which she was pursuing her education and research, deeply intertwined with her identity and her diversity group. The reality is that certain individuals are metaphorically on the ``Raft of Medusa'', barely staying afloat, while others are on comfortable yachts or even massive battleships. 



\section{On ECCV 2022, Tel Aviv, Israel}
\label{sec:israel}
\epigraph{\textbf{TL;DR. We used to be human beings! But not in the sense of dehumanization.}}


\begin{figure}
     \centering
     \begin{subfigure}[b]{0.45\textwidth}
         \centering
         \includegraphics[width=\textwidth]{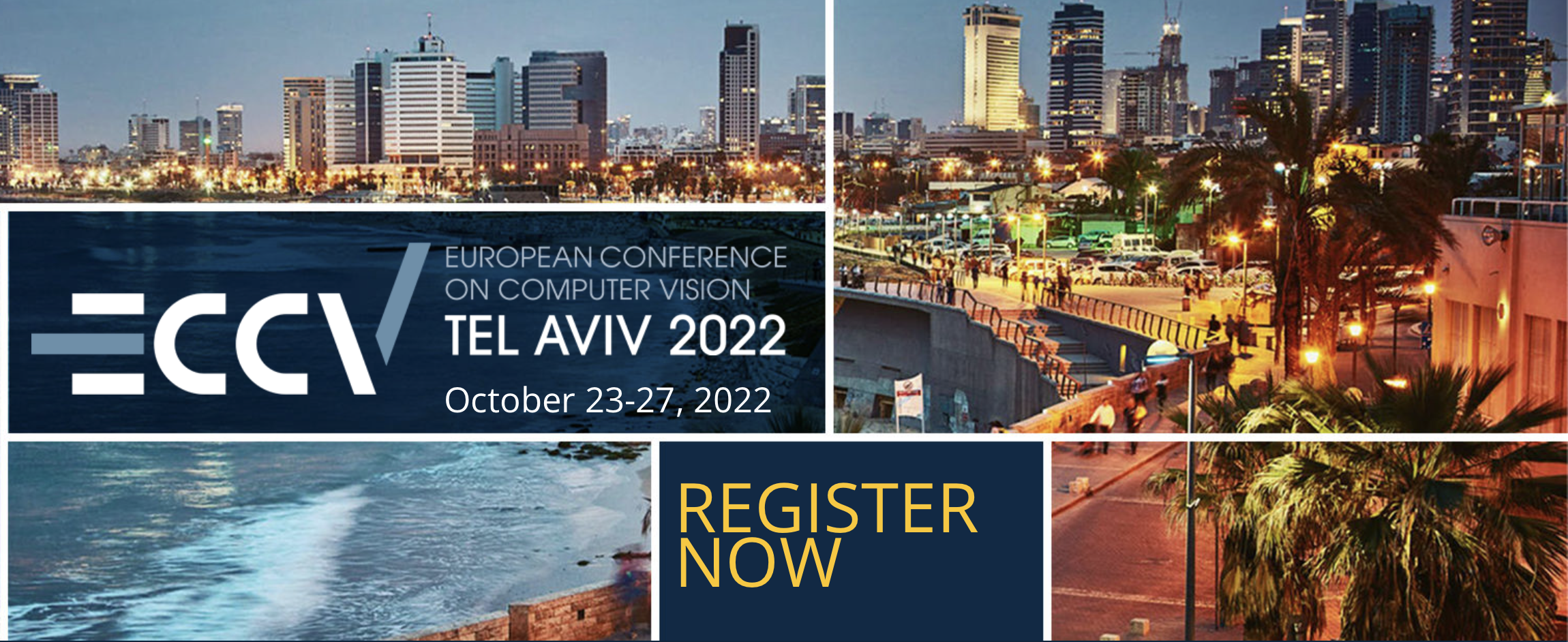}
         \caption{}
     \end{subfigure}
     
     \begin{subfigure}[b]{0.45\textwidth}
         \centering
         \includegraphics[width=\textwidth]{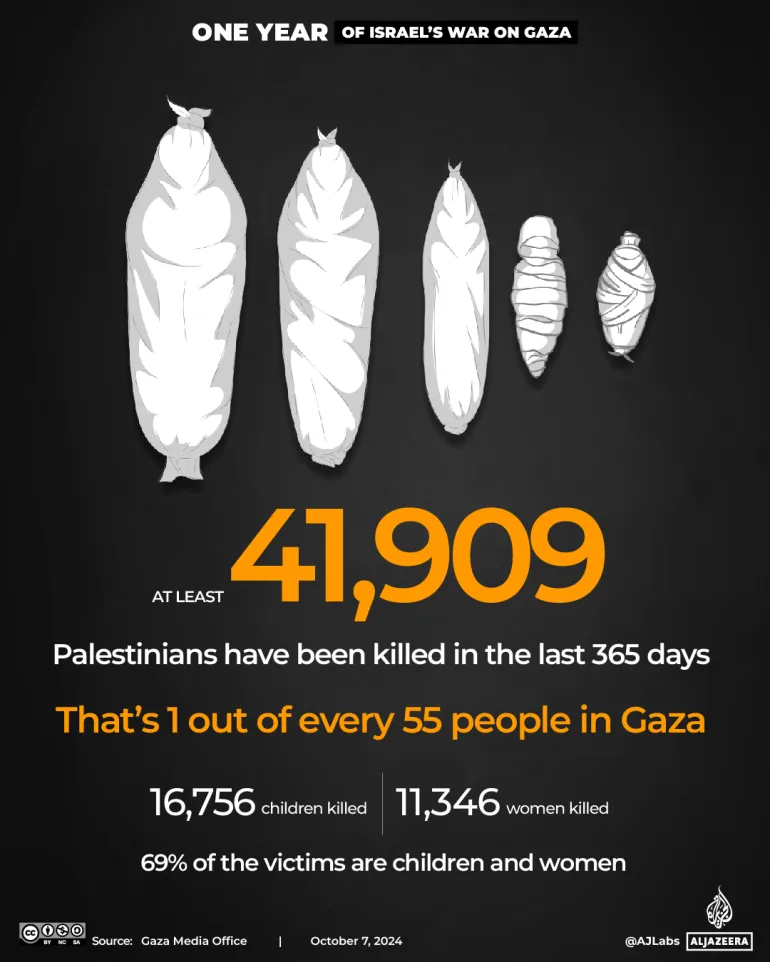}
         \caption{}
     \end{subfigure}
\caption{It provides evidence on how it was normalized to situate ECCV in the place of a long history of human rights violations by the Israeli government which is still ongoing. (a) Oct. 2022, ECCV, Image taken from~\cite{eccv}. (b) Oct. 2024, Image taken from~\cite{gaza2}. An analogy to this would be holding a conference in the backyard of the holocaust while it was happening, and then inviting Jewish participants to that conference!}
\vspace{-1em}
\label{fig:Israel}
\end{figure}



Quoting the UN's special rapporteur on Palestine ``Israel's Genocide of Palestinians in Gaza is an escalatory stage of a long-standing settler colonial process of erasure''~\cite{gaza1}. Over the past year, Israeli attacks have killed at least 41,909 Palestinians living in Gaza, equal to 1 out of every 55 people living there, refer to Fig.~\ref{fig:Israel}. ``At least 16,756 children have been killed, the highest number of children recorded in a single year of conflict over the past two decades. More than 17,000 children have lost one or both parents.''~\cite{gaza2}. Looking at a humanistic view through the lens of a talented artist on the situation in Gaza~\cite{gaza3}, we see dreadful war crimes that no one in our community dares to speak about. We were once human beings, and not in the sense of dehumanizing or depriving anyone of their rights, but spiritually, we have lost our humanity by turning away and watching these atrocities unfold.

What does it mean to call for responsible AI when these violations are not openly advocated against in our community? What does it mean to work on trustworthy ML when researchers organizing major conferences fail to condemn this? These are undeniable facts. 
The very fact that a major conference like ECCV was held in Israel, normalizes these crimes, Fig.~\ref{fig:Israel}. Not only that, but we are actively excluding those who speak out for these causes. This is deeply personal for most Muslims and/or Arabs including our researcher, the ``devil'', especially taken into account she has witnessed firsthand what has been done to the children in Gaza. Referring to the Israeli historian, Ilan Pappe, he has defended the right of Palestinians to return to their homeland~\cite{chomsky2015palestine}. So, what does it mean to celebrate and host such a conference in a place where Palestinians, who were born and used to live there, are barred from returning? 


\begin{figure*}[t]
\vspace{-1em}
     \centering
     \begin{subfigure}[b]{0.24\textwidth}
         \centering
         \includegraphics[width=\textwidth]{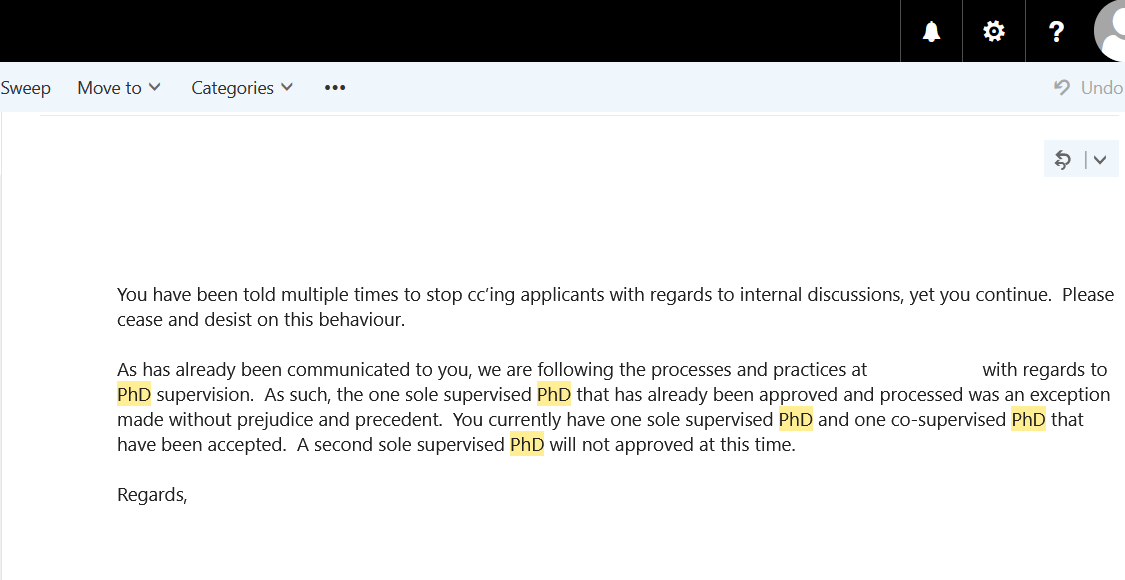}
         \caption{}
     \end{subfigure}%
     \begin{subfigure}[b]{0.24\textwidth}
         \centering
         \includegraphics[width=\textwidth]{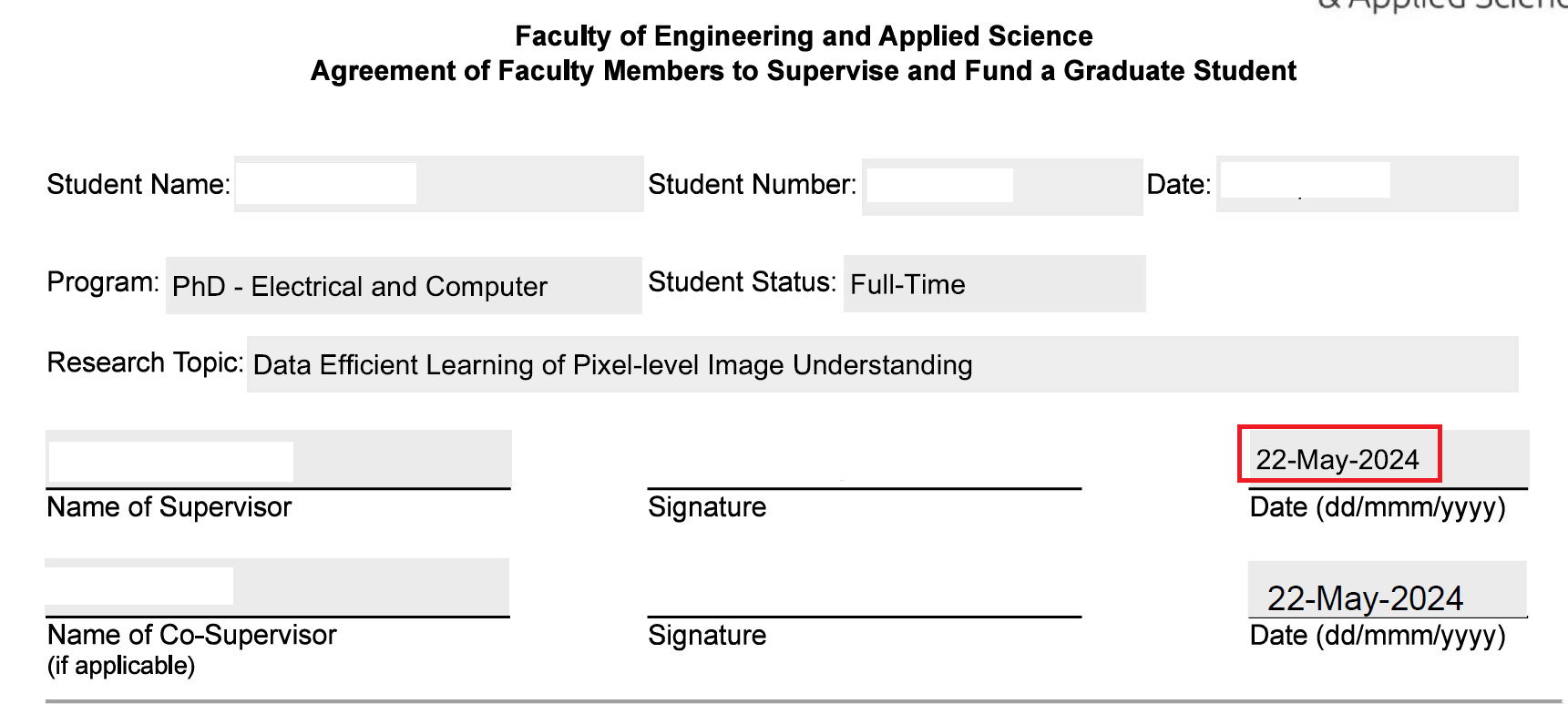}
         \caption{}
     \end{subfigure}%
     \begin{subfigure}[b]{0.24\textwidth}
         \centering
         \includegraphics[width=\textwidth]{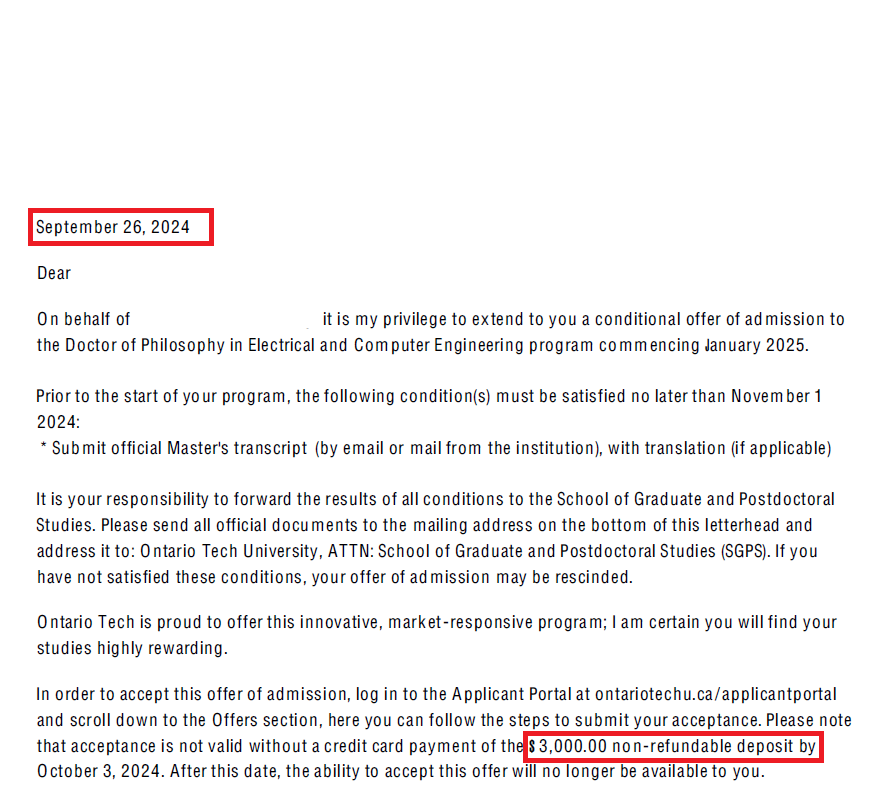}
         \caption{}
     \end{subfigure}
     \begin{subfigure}[b]{0.24\textwidth}
         \centering
         \includegraphics[width=\textwidth]{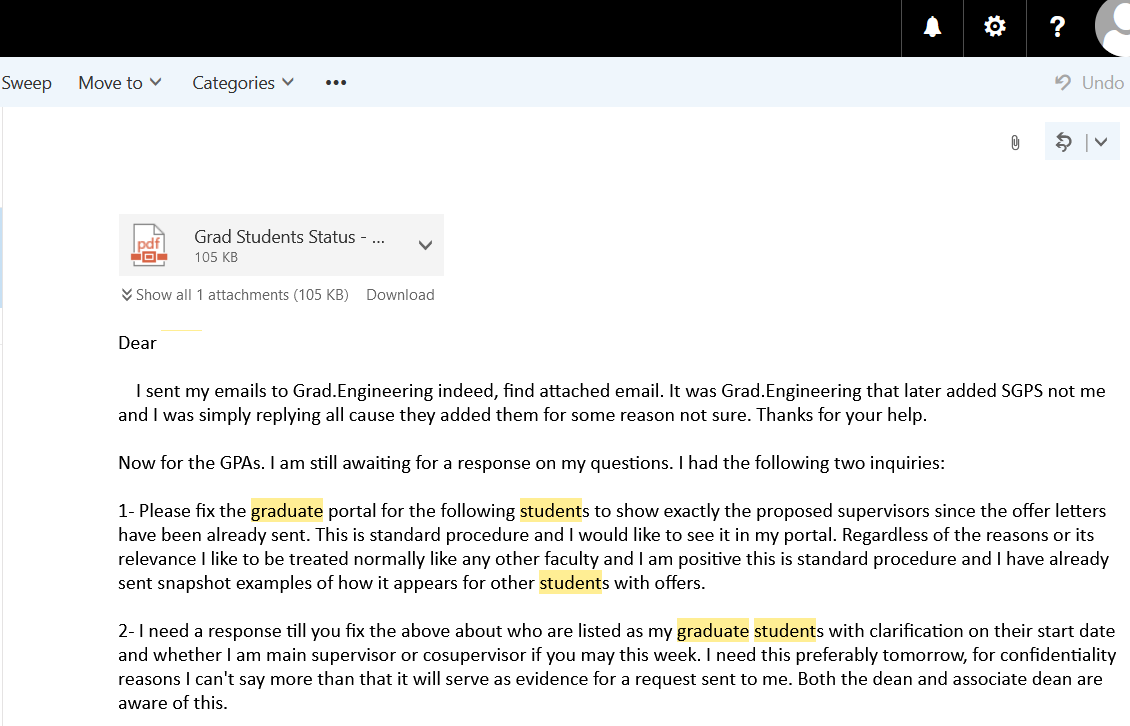}
         \caption{}
     \end{subfigure}

     \begin{subfigure}[b]{0.19\textwidth}
         \centering
         \includegraphics[width=\textwidth]{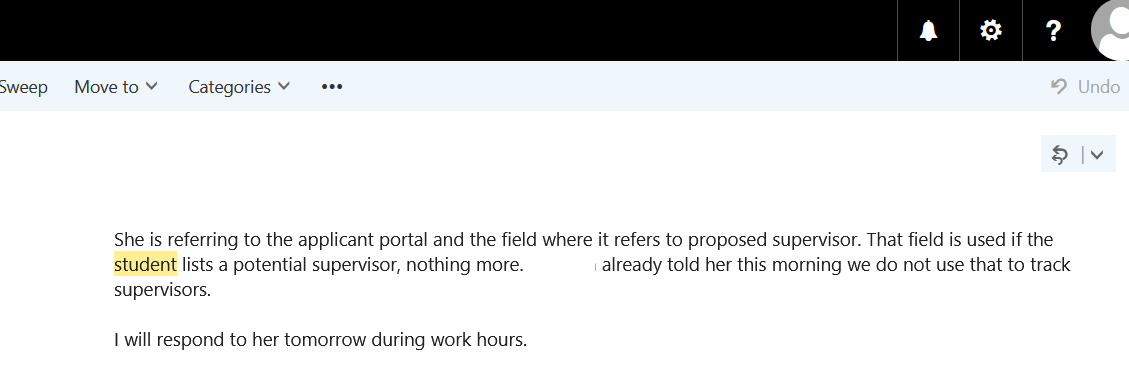}
         \caption{}
     \end{subfigure}%
     \begin{subfigure}[b]{0.19\textwidth}
         \centering
         \includegraphics[width=\textwidth]{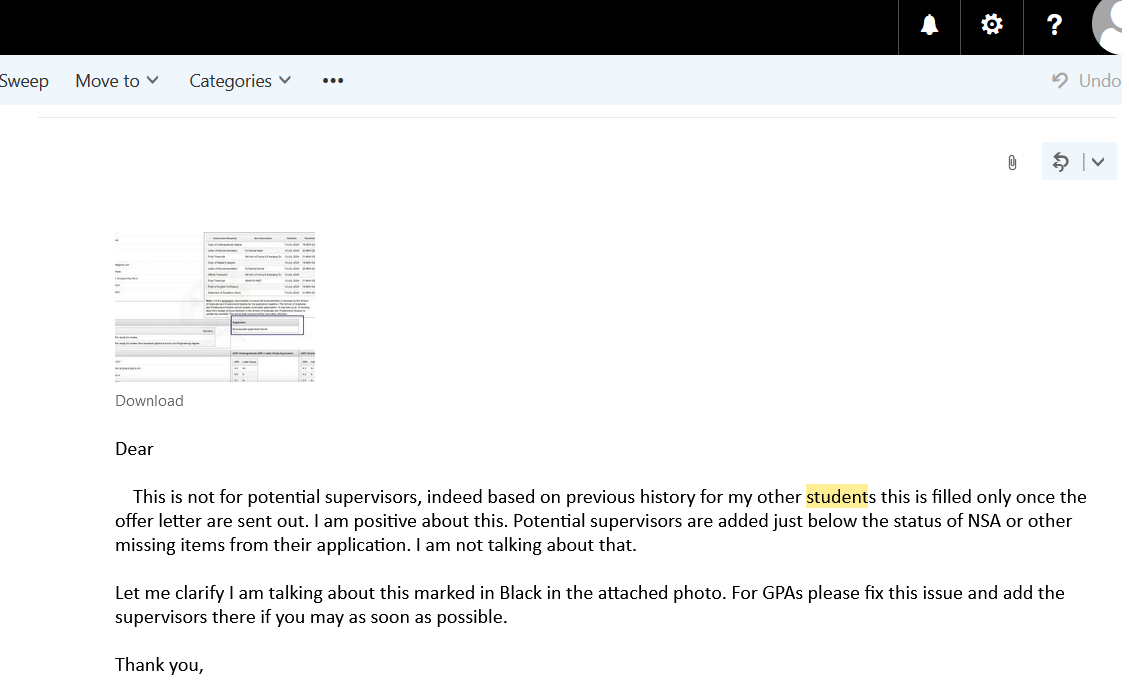}
         \caption{}
     \end{subfigure}
     \begin{subfigure}[b]{0.19\textwidth}
         \centering
         \includegraphics[width=\textwidth]{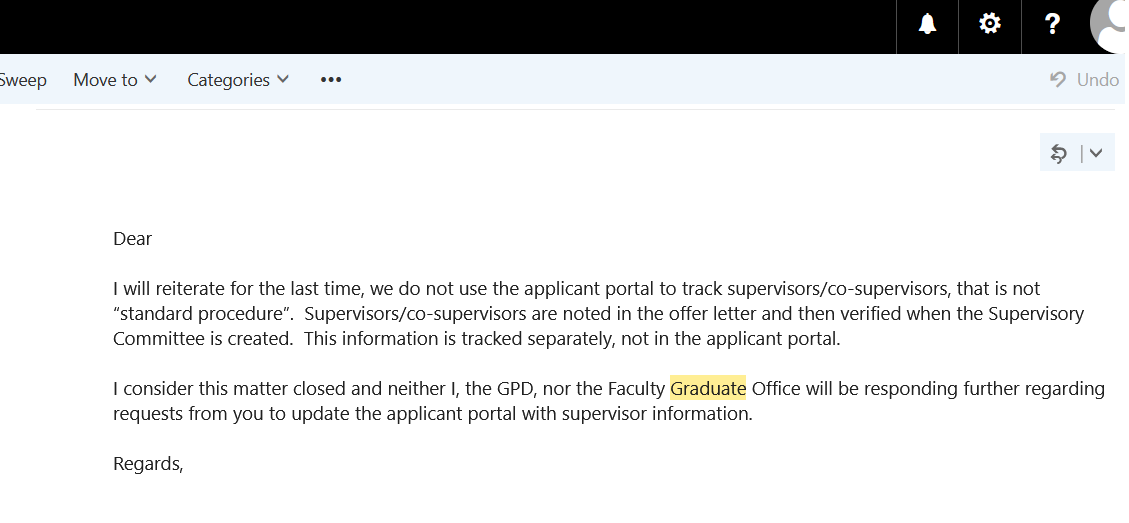}
         \caption{}
     \end{subfigure}%
     \begin{subfigure}[b]{0.19\textwidth}
         \centering
         \includegraphics[width=\textwidth]{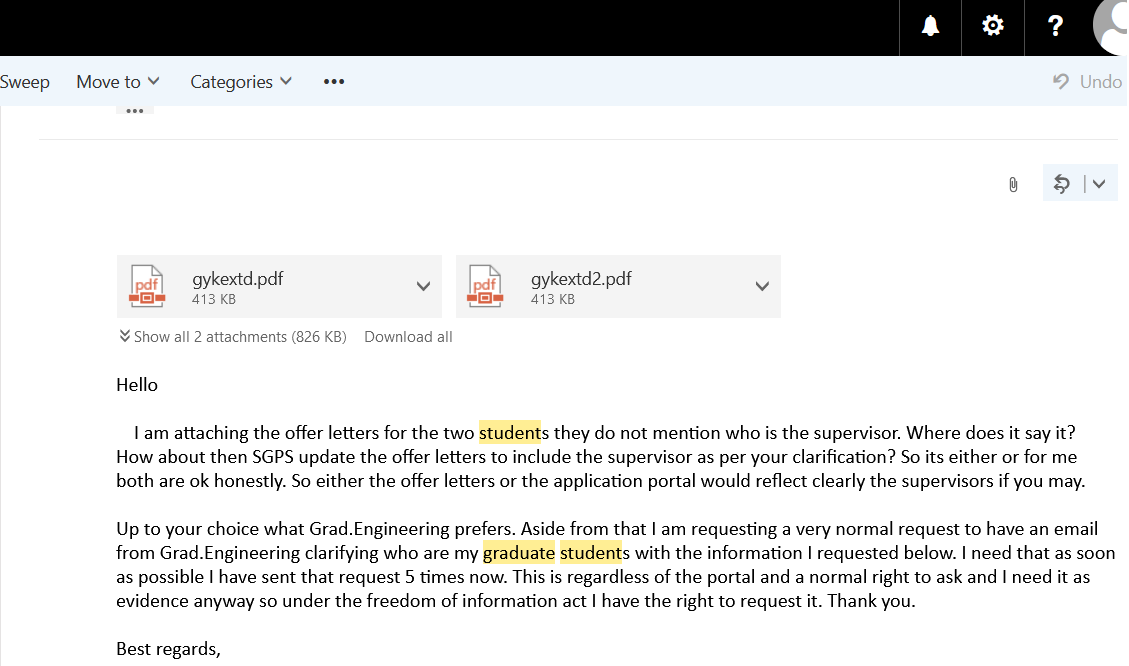}
         \caption{}
     \end{subfigure}%
     \begin{subfigure}[b]{0.19\textwidth}
         \centering
         \includegraphics[width=\textwidth]{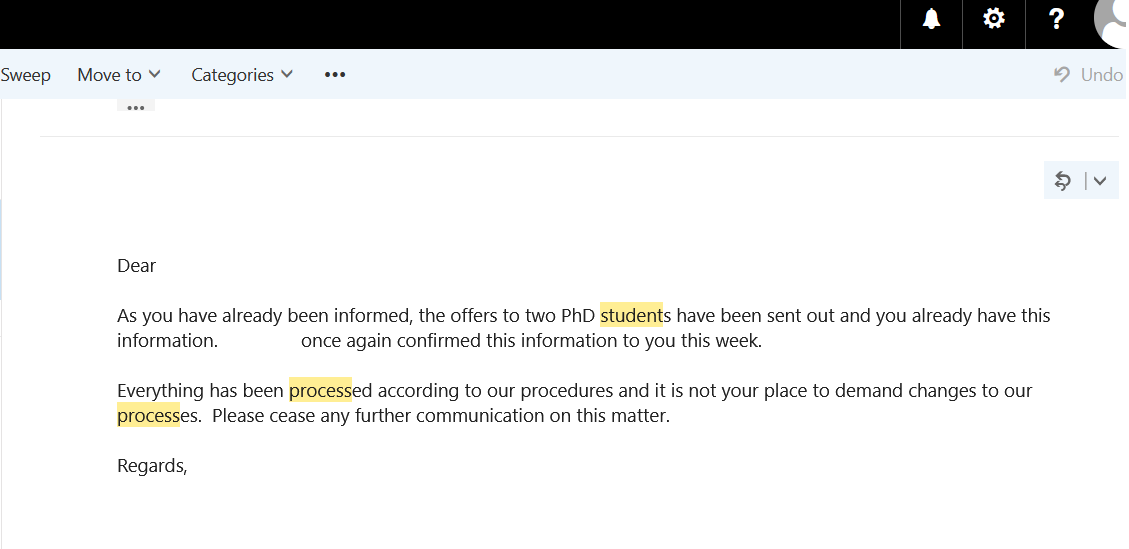}
         \caption{}
     \end{subfigure}
\caption{A year and a half with grants and two affiliations but still zero students. (a) Being prevented from admitting a sole supervised second student. Except, the graduate program rejected the deferral and rescinded the offer of her first student due to visa delays, and her co-supervised student ended up to work on a completely different topic that does not intersect with computer vision. (b, c) From 22 May a request sent to admit a student, till 26 Sep. when the actual offer letter was sent. (c) 3,000\$ admission fees instead of the standard 100\$, that were waived after fighting for these international students rights. Resulting in zero students and her inability to start the lab. (d-i) Being denied access to standard information about the students that are supposedly joining the devil's lab in Jan 2025. It can be indicative that they were never registered under her supervision, which could have resulted in another year or two without students. At this point there was nothing to prevent the university from this behaviour, not even Grievances worked.}
\label{fig:nostudents}
\end{figure*}

While there is an ill-posed argument that the Israeli government is defending her citizens, war crimes should never be justified. Nothing justifies the killing of civilians especially children and violating the major human right to life and liberty. What safe AI are we pursuing if we think that it is permissible to kill all these children under any circumstance, not even a single one, not even being kidnapped. If we normalize this now as researchers in AI, do we really believe any safe outcome will emerge out of this later. This is an invitation to everyone who has the courage to speak out against these violations in our community to invite the organizers to put on their websites that they are against this Genocide. If this happened for the next five years in all our top conferences, i.e., CVPR, ICCV, ECCV, NeurIPS, ICML, ICLR, ICRA, IROS, AAAI, and IJCAI, only then we can guarantee there is no discrimination against any of the researchers in the field who share the same sentiment. State it out loud and let everyone know, including our sponsors, that our computer vision community is against this Genocide. If the community was not able to do it, then it is implicitly indicative of the amount of power such a government has on our community and the pressure researchers who are voicing concerns on their violations are facing. At the end, Palestinians like everyone else have the right to life and liberty especially the children so this should not be debated.




\section{The Faculty Experience}
\label{sec:faculty}
\epigraph{\textbf{TL;DR. A faculty was prevented from starting her research lab violating academic freedom!}}


\subsection{The Faculty Phase}
In this section, we discuss the devil's experience as faculty in Canada. In summary it included; zero university offers that had computer vision labs, zero students, no control over grant money, three xxxxxxxxxxxxx and harrasment.

\textbf{Faculty hiring.}  First, let's refer to the statistics on the proportion of Muslims in the Greater Toronto Area (GTA) and Montreal which were at 10\% and 8.7\%~\cite{CANStats}, respectively. Now referring to all the faculty in computer vision across Canada, i.e., faculty that are actively publishing in top computer vision conferences, to the best of our knowledge not a single one of them is a Hijabi, which is a visible minority. That does not necessarily indicate discrimination, it could potentially be that Muslim Hijabi women are not interested in this field or not as good in it in Canada. Except the devil, who happens to be African Canadian Hijabi, had ten top quality works in either A/A* conferences or journals with high impact factor, and some of them are even under Responsible AI and vision for robotics. So if a researcher like her was denied the opportunity to open her research lab without a reason, that does violate her right to contribute to the research community. It is even worse considering that Canadian laws in hiring faculty, prioritize Canadians and permanent residents over foreign workers. Second, the devil applied to the following universities and was not even invited for an interview, even when she was applying with grants acquired; Simon Frasier University, York University, University of Toronto, University of Waterloo and University of Guelph. In fact she applied in various departments within these universities, yet she was told she was not sufficiently good to be part of these universities. That could be understandable to some aspect, it is true the majority of the faculty hired there have better profiles, but there are others that are not. Let's assume that the quality of faculty profiles could be relative, with other factors coming in play. She did not apply outside Ontario and British Columbia to avoid discrimination that is quite pervasive in other provinces to her diversity group, so taken that into account she accepted an offer in a small university in Ontario. 

\begin{figure}
     \centering
     \includegraphics[width=0.5\textwidth]{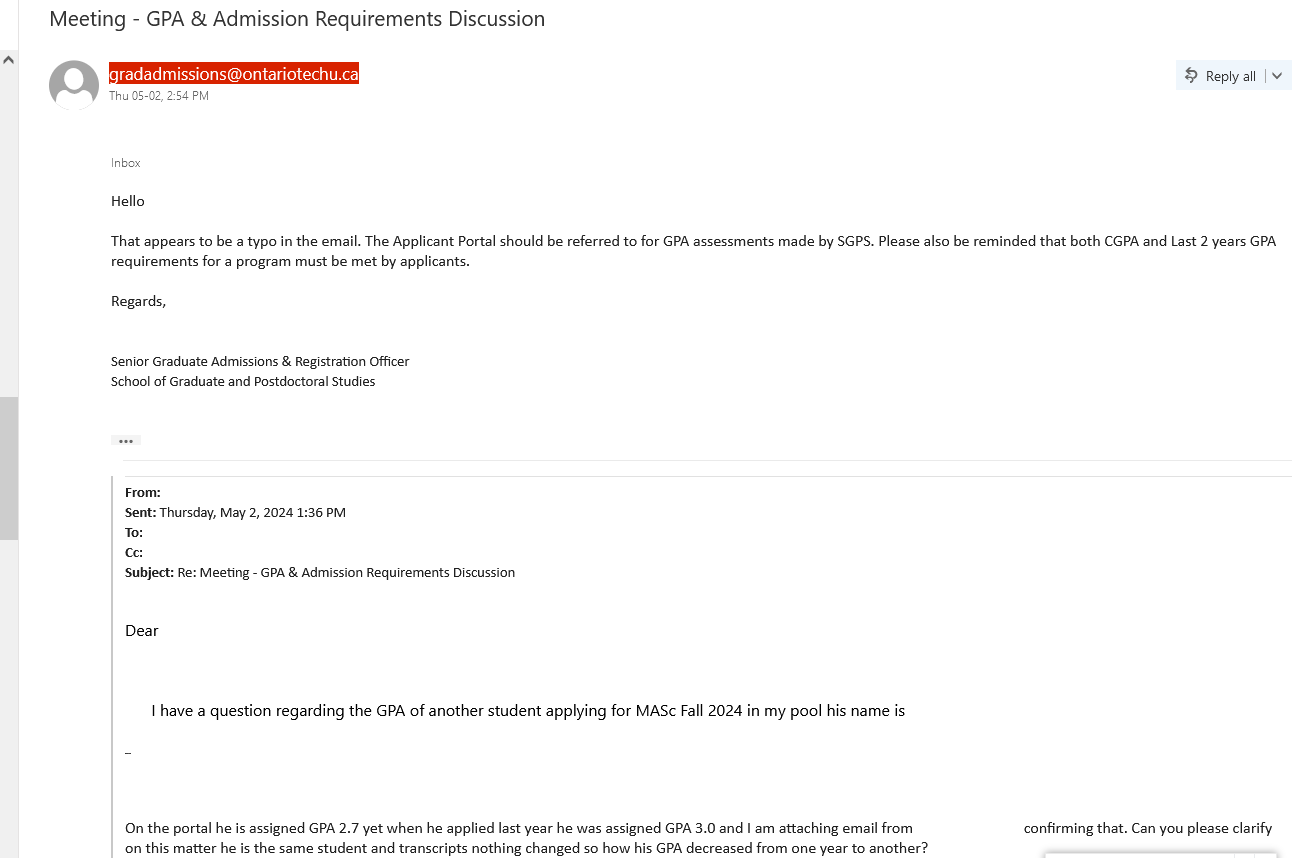}
     \caption{GPA of one Black African MASc applicant changing from one year at 3.0 to another year at 2.7 without a change to his transcripts and referred to as a typo.}
     \label{fig:gpa}
\end{figure}

\begin{figure}
     \centering
         \includegraphics[width=0.49\textwidth]{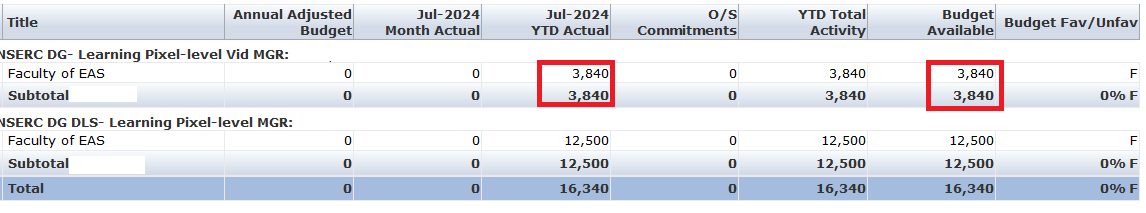}
     
\caption{Receiving only 3-4K\$ from 24K\$ yearly installments, resulting in delayed students admission.}
\label{fig:grant}
\vspace{-1em}
\end{figure}


\textbf{Faculty support system.} As any faculty starting her lab first year she admitted a MASc student, whose deferral for visa delays was rejected by the graduate program director not allowing her to continue her supervision for no real reason\footnote{\url{https://www.dropbox.com/scl/fo/j1yma0y2nzd5fwjj3oy3w/AInO0Tksc2OYg4bzzxxlz-I?rlkey=mn71egnyvk0skn48u06ntyyom&st=fst26f21&dl=0}}. Not only that but even after acquiring grants and requesting to admit PhD students in her second year, she was told that she is not qualified and does not have the necessary experience to supervise solely PhD students. Referring to this university policy that states ``Be the sole supervisor of a PhD thesis student. This is restricted to Graduate Faculty who have advanced experience as appropriate for the graduate program''\footnote{\url{https://usgc.ontariotechu.ca/policy/policy-library/policies/academic/graduate-faculty-appointments.php}}. After multiple negotiations, they made an exception for one student but prevented her from a second student, refer to Fig.~\ref{fig:nostudents}(a)\footnote{\url{https://www.dropbox.com/scl/fo/qy3yidh5veohvp2i7evze/AECFAP8th3lnM_ztyBdE9BQ?rlkey=l5kg61zfa0iycyl7ss8v4yla4&st=kc4b425o&dl=0}}. The first student for visa delays requested deferral that was rejected by the graduate program, yet again\footnote{\url{https://www.dropbox.com/scl/fi/nzbdyi7wv0o6to123sr3w/6-FirstStudentDeferral-1.pdf?rlkey=1g5ngbpwi7afj2zs6litrfec9&st=taix1z2r&dl=0}}. In a small university with a department that had no other computer vision faculty, it ended up in all refusing to co-supervise her students. Instead, she tried to co-supervise a PhD student with another faculty, the student was initially applying with her which she shared with that faculty. But the student ended up working on topics far from computer vision diverging from the original plan as such she requested to stop co-supervising him. 

She tried to collaborate with other researchers outside the department. But then new problems in students admission appeared. After finding a good faculty to co-supervise her student that passed admission requirements, a request was sent for his admission on May 22, yet he received the offer on 26 September after four months. Furthermore, the university was requesting 2,000-3,000\$ admission fees instead of the standard 100\$, refer to Fig~\ref{fig:nostudents}(b-c)\footnote{\url{https://www.dropbox.com/scl/fo/1z5kkbfpi9rneyflm7r5f/ADBvEMpZ5nAGozq8frZl4lo?rlkey=23blp9jeu63d1ndwdjxhp0jxb&st=l9mnkowh&dl=0}}. After multiple discussions to fight for these international students rights to be admitted as standard, they waived these fees for them as fully funded students. After long negotiations, one offer letter in addition to the previous one were sent to two PhD students to join Jan 2025 her lab, except they were not identified as her students anywhere. Neither the portal nor offer letters not even through email requests as standard, refer to Fig.~\ref{fig:nostudents} (d-i). She was threatened to seize from inquiries to identify who are the registered students under her supervision\footnote{\url{https://www.dropbox.com/scl/fi/ig9hzn0jldzvqc72y83g4/InquiryAboutStudents-1.pdf?rlkey=qx1rok7o02b6eiy8mq7lumuop&st=3sfxis7v&dl=0}}! This can be an indicative that these students were registered under the supervision of another faculty, and that the devil might have stayed for another year or two without students. The only student she was allowed to supervise was a Master's of Engineering student working on a project/report, in addition to one visiting PhD student who visited in summer 2023 and 2024. Grievances did not help, so the only way out for that faculty was to simply leave. Inquiring from other faculty colleagues outside the university, they had two MSc students in the first year and two PhD joining in the second. Hence, it is quite standard in our field and it is quite unexpected for the devil to have zero students after a year and a half when her university has three admission cycles as well during Fall, Winter and Spring.

Other incidents regarding her students admission include a Black African MASc student that got evaluated in one year 3.0 for his GPA, at that point the devil was told that the minimum requirements is at 3.3. The next year same student applied with same transcripts in the hope for a second shot with that faculty and now that she knew the minimum requirements for MASc GPA is actually at 3.0 not 3.3. Except this time his GPA was evaluated to be 2.7. The admission staff referred to this incident as a typo, refer to Fig.~\ref{fig:gpa}\footnote{\url{https://www.dropbox.com/scl/fi/zdm7athfrzkmn4py8vgym/1-GPAS-1.pdf?rlkey=lp4swu5eb4t44jaazhkf2xc3v&st=6y0t2wcy&dl=0}}, since the first time the GPA was sent through emails not on the portal as the faculty did not have access then to it for administrative reasons. Except normally there should never be a typo for student GPAs, since it is the difference between an accept or reject for such under-represented groups.

Now for the grant money that she was never really able to use. Her discovery grant that was valued at 24,000\$ per year over five years but ended up to be at around 4,000\$ in her research account, Fig.~\ref{fig:grant}. She was told this was standard procedure in the university related to receiving the money from the funding agency as installments over the year. Except, the funding agency did not have that in writing in any of the grant documents and their payment schedule for research support fund states otherwise\footnote{\url{https://www.rsf-fsr.gc.ca/administer-administrer/payment-paiement-eng.aspx}}. Nonetheless, she was not sure if these grants are considered under research support fund so she left it at that, but these installments within the yearly installment does impact the ability of young researchers to admit students and could possibly be one of the reasons behind such delayed students admission. 

\begin{figure}
     \centering
         \includegraphics[width=0.49\textwidth]{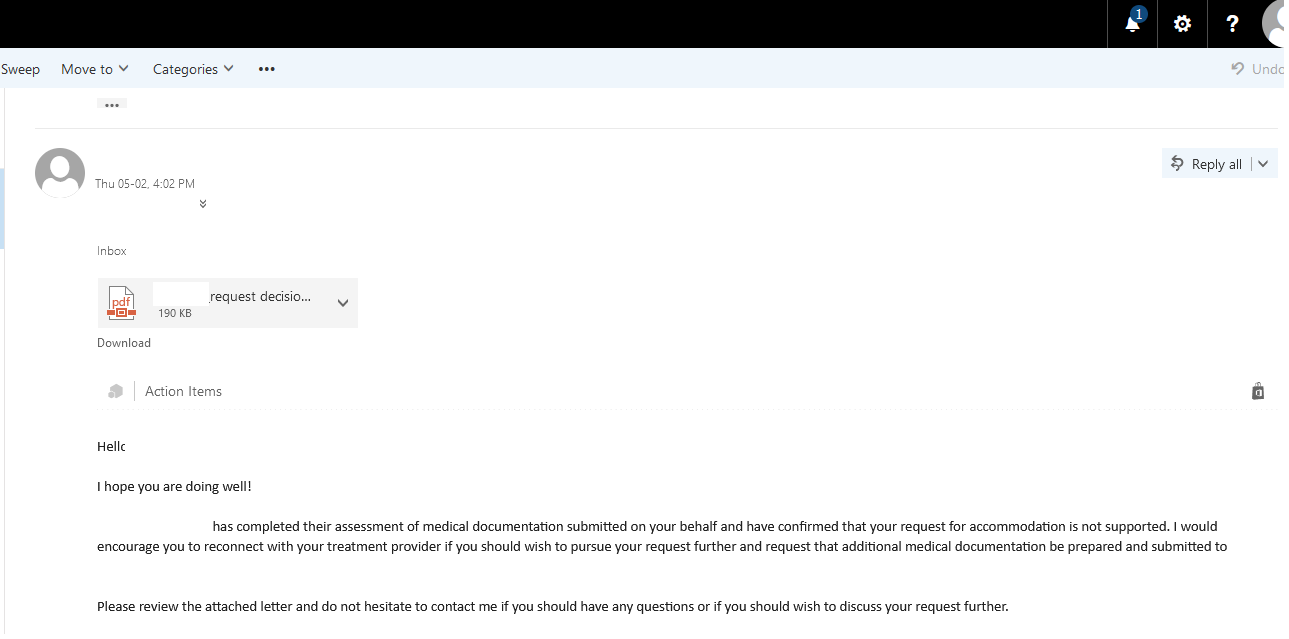}
     
\caption{Accommodation rejected for a tenure-track not teaching faculty to teach two days a week because of her plantar fasciitis.}
\vspace{-1em}
\label{fig:humiliation}
\end{figure}

Aside from the issues in establishing her lab, the devil was humiliated in her university through various forms. She was denied a simple accommodation to teach two days a week because of plantar fasciitis that affected her feet with a confirmed case, Fig.~\ref{fig:humiliation}, this had relation to her one hour drive to the university. Yet she was told she had to teach three days a week even when she is not a teaching faculty and had such a health condition. Her university insisted that she teaches her course in Ramadan during breaking her fast on the sunset. Due to her fear of retaliation she did not complaint taken into account that she was already complaining about her students admission. But she had to teach while eating/drinking a bit so she does not lose strength from not eating or drinking all day. Full context provided\footnote{\url{https://www.dropbox.com/scl/fo/mruamheqgq4jumpfwswu2/ANCaqdVsR-JDocHgN3BO57M?rlkey=6szmbe7q386bdp68744v1lss4&st=6nmb5hq6&dl=0}}, where the goal was to demean her in front of students and staff. 

In summary, she was going through implicit constructive dismissal. While the devil tried to reach out to Women in Machine Learning, Black in AI and Deep Learning Indaba for legal and financial advice, she did not receive a response. 



\begin{figure}
     \centering
         \includegraphics[width=0.49\textwidth]{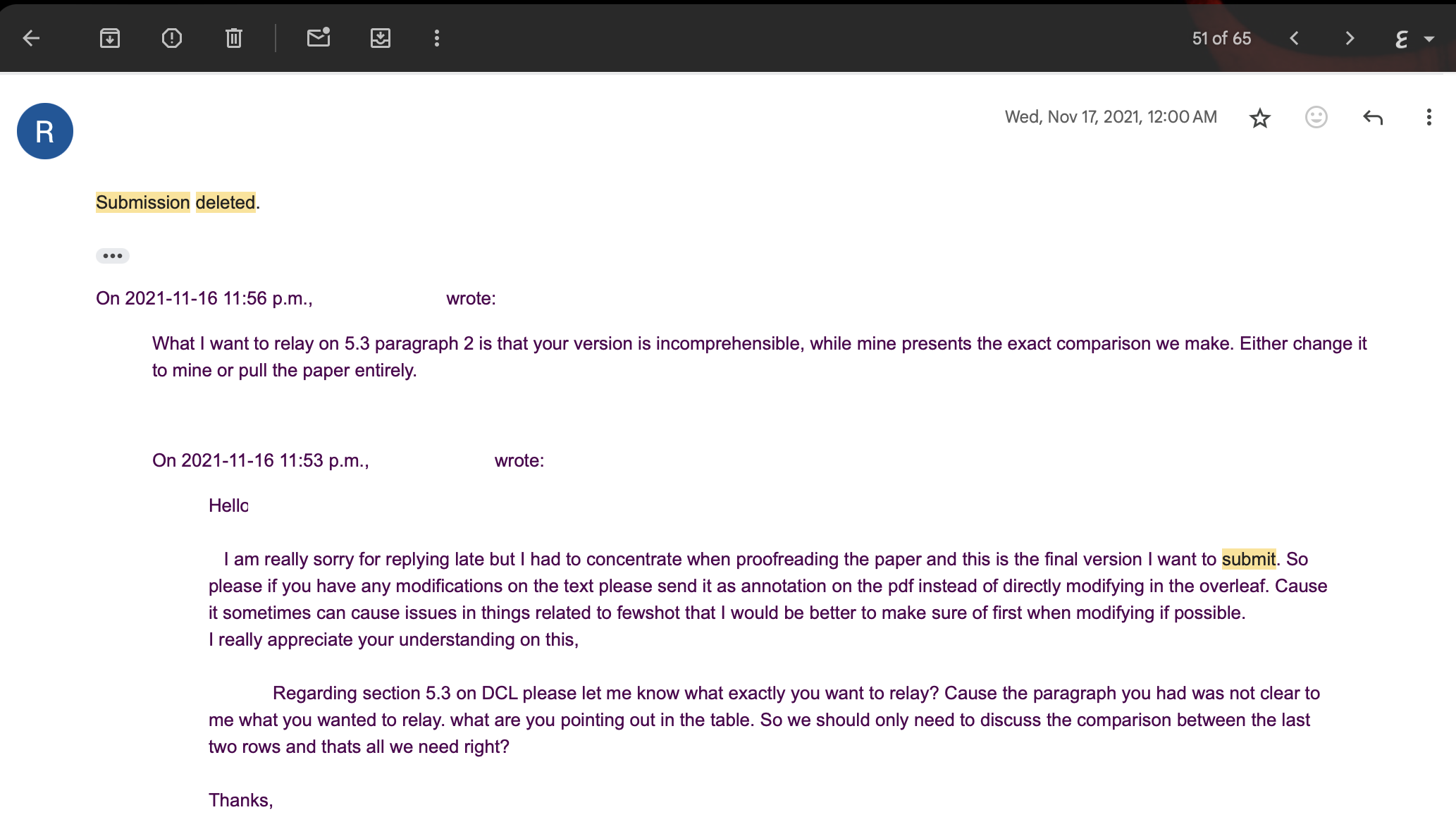}
     
\caption{The incident during the devil's postdoc phase.}
\label{fig:Richard}
\vspace{-1em}
\end{figure}

\subsection{The Pre-Faculty, Postdoc Phase}

From the previous, the question on why she ended up in this university from the first place emerges. So let's go back to the phase before it, during her postdoctoral fellowship. Through this phase there were lots of hazing that she was not good enough for academia. She was even implicitly told she was a nobody in the field by her supervisors and she was never really promoted for her research by the university. Yet, it was still relatively successful postdoc with three high quality publications in CVPR and TPAMI and three publications that are being reviewed. For her diversity group it was good achievements, but when looking to other researchers in the field she was quite behind. She barely could publish one high quality publication per year while others at that stage could publish 2-3 publications or more. This is not for the sake of comparison but rather to clarify that the field is indeed competitive. 

She was temporarily affiliated with quite a famous Canadian institute, except that was taken away even before her postdoc time finished and she was not able to access the computational resources she initially had. In fact she was later denied an affiliation as faculty with that institution where multiple times the staff avoided her question of when they will open the application. All of this started from a reprisal case because she requested from one of her supervisors to be listed as equally advising on one of the works that she mentored his PhD student. Since most of the ideas were hers and third of the work was her implementation. That was denied and it sufficiently angered her supervisor to prevent her from continuing another project with same student after five months of readings and meetings. 

\begin{figure*}[t]
     \centering
     \includegraphics[width=\textwidth]{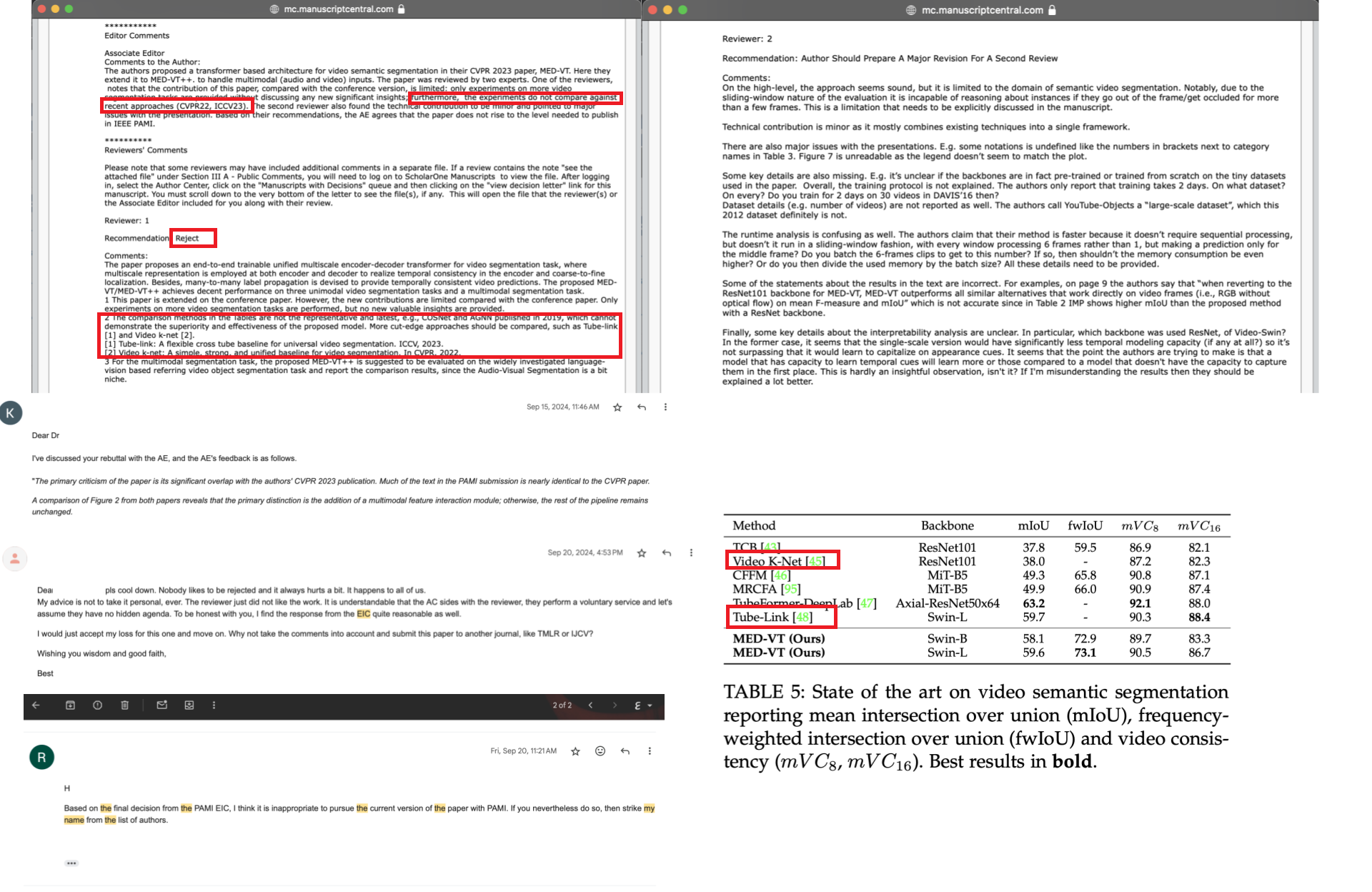}
     \vspace{-3em}
     \caption{TPAMI standard reviewing process that included a false claim for not comparing to two works that were indeed compared against. Yet the editorial board rejected the appeal. It also resulted in potentially stopping the project if it were not for the devil interfering to continue it herself!}
     \label{fig:tpami_reviewing}
\end{figure*}

Her other supervisor offered her to be the last author and to guide a different PhD student work, since he was retiring. She accepted and was quite grateful for it, especially that it was accepted in CVPR. However, when the extension of the same work in TPAMI got rejected in an unfair way, her supervisor requested to be removed from the authorship and his student stopped from doing any work on the re-submission. So after a year and a half of research efforts and meetings, they were suggesting to stop that project, yet again. Refer to Sec.~\ref{sec:reviewing}. Her supervisor was promoting that ``She is hard to work with''. Except, in one of her submissions to CVPR, her supervisor deleted the submission one night before the deadline and after the deadline for abstract submission. The reason for that was a dispute on a minor paragraph rewrite, refer to Fig.~\ref{fig:Richard}. Her supervisor's wife encouraged her to open her research lab ``back'' in her country. While she did this with good intentions, except our devil is Canadian as well so why not in Canada! 

\section{On Our Not-so-Broken Reviewing System}
\label{sec:reviewing}

\begin{figure}[t]
     \centering
     \includegraphics[width=0.5\textwidth]{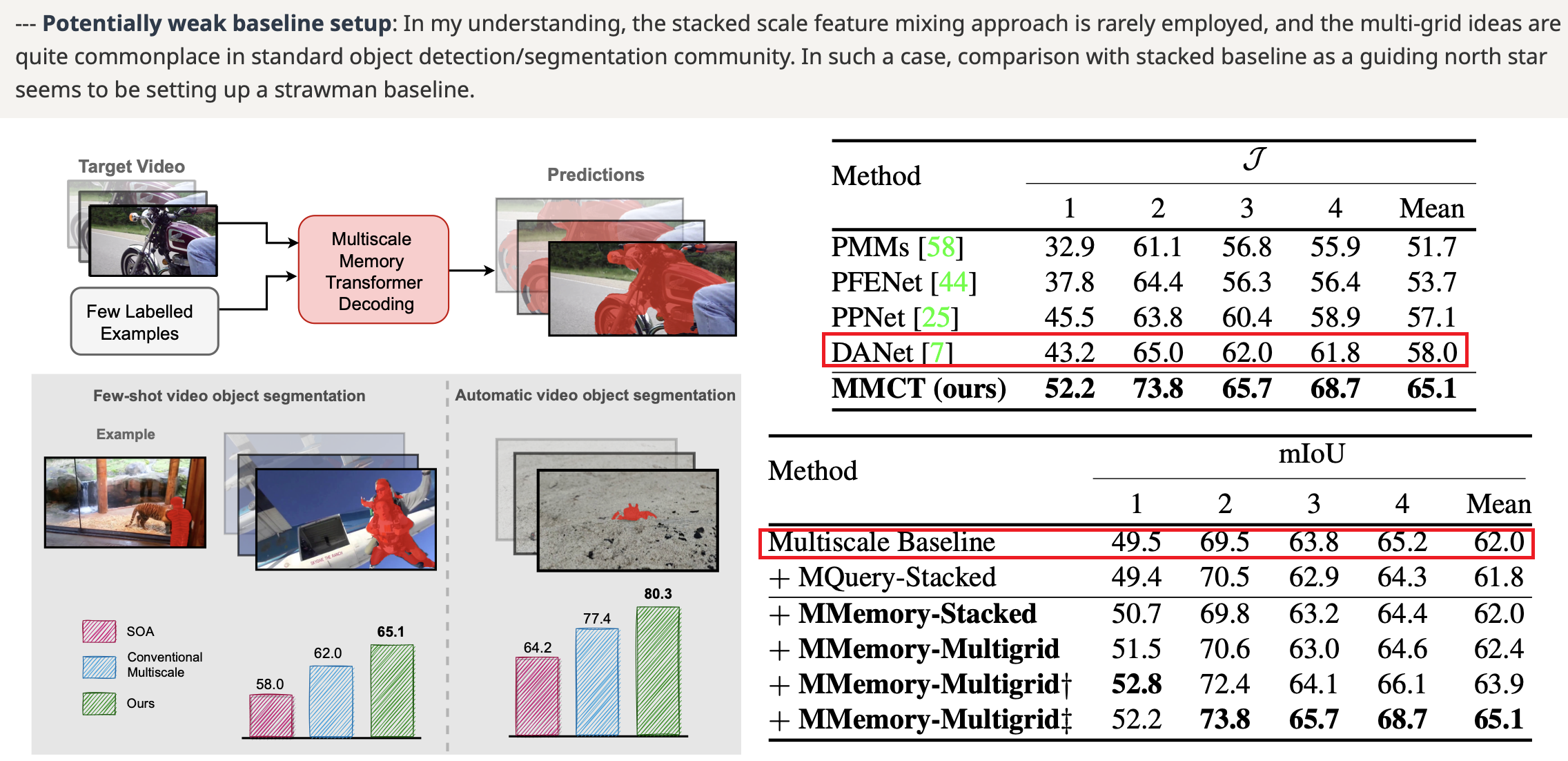}
     \caption{CVPR 2023 rejection pointing to a weak baseline setup when the baseline is outperforming the state-of-the-art and adapting a recent method. Same paper got rejected in Neurips 2022 and ICCV 2023.}
     \label{fig:cvpr23}
     \vspace{-2em}
\end{figure}

\begin{figure*}[t]
     \centering
     \includegraphics[width=0.75\textwidth]{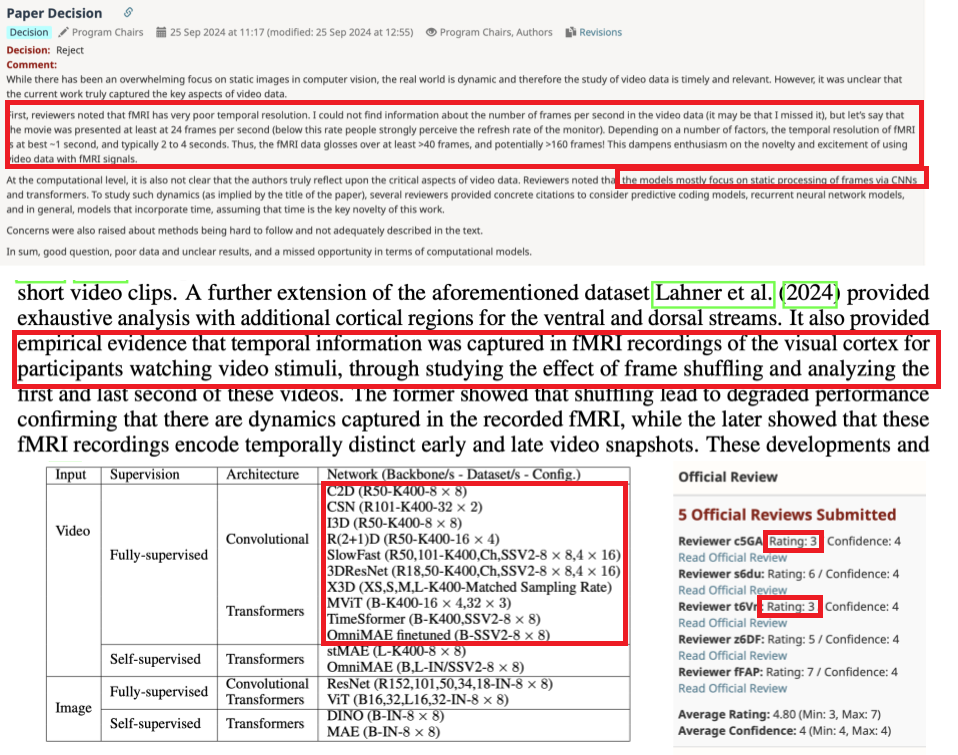}
     \caption{NeurIPS rejection killing the research direction with unfounded claim that there is dampened enthusiasm in studying video stimuli with fMRI recordings when recent research published in Nature Communications shows otherwise.}
     \label{fig:neurips2024}
     \vspace{-1em}
\end{figure*}

\epigraph{\textbf{TL;DR. We used to be critical thinkers until we became greedy.}}

Various discussions occurred in CVPR 2023 and 2024 regarding whether the reviewing system is broken or not. There is an important clarification that this is an existential problem, meaning if there exists one researcher with strong evidence showing obvious flaws in the reviewing system then there is a problem. Whether it is a universal problem or not is insignificant, this is with reference to (UDHR Art. 27~\cite{prabhakaran2022human}). To frame it with a strong analogy, if someone got killed inhumanely without any reason, it does not make sense to bring everyone else to talk about how they are living in luxury and enjoying their lives. That still does not deny the previous person from his right to live! Nonetheless, quoting~\cite{su2021affective} ``While many describe excitement and success,'' in Computer Vision, ``we found strikingly frequent feelings of isolation, cynicism, apathy, and exasperation over the state of the field. This is especially true among people who do not share the unbridled enthusiasm for normative standards for computer vision research and who do not see themselves as part of the in-crowd.'' In this section, we document the devil's experience in the reviewing process to what could either be ``normative standards'' or the result of a broken system. 

The devil's TPAMI submission~\footnote{\url{https://arxiv.org/pdf/2304.05930}} that is a CVPR extension~\footnote{\url{https://openaccess.thecvf.com/content/CVPR2023/papers/Karim_MED-VT_Multiscale_Encoder-Decoder_Video_Transformer_With_Application_To_Object_Segmentation_CVPR_2023_paper.pdf}}, which she was the last author and excitedly her first evidence of research work coming out of her lab, was rejected for false claims. The decision was not ``Revise and resubmit as new'', it was ``Reject''. Two major reasons include how the work's contribution was not sufficient and that the authors were not comparing to state-of-the-art methods with two references cited. Except, these two references were already compared to in the work and they were extending the paper in three major ways. While one reason could be subjective the other was a clear false claim and major flaw in the review. Thus, the devil sent to two researchers in the editorial board to request an appeal. They responded that they followed the standard reviewing procedures and referring to only one of the reasons, Fig.~\ref{fig:tpami_reviewing}. Except, the devil requested another reviewer instead of that reviewer with obvious false claim, especially given that the second reviewer gave a ``Major revision'' decision. Even with these obvious flaws in the review they still denied that appeal and citing that researchers should not take these rejections personally. Yet, it is these rejections with obvious flaws that do affect researchers career and ended up that the devil getting only an appointment in a small university, limiting her grants and so forth. Another TPAMI submission\footnote{\url{https://arxiv.org/abs/2211.01783}} received the first round of reviews after one year and a half, as such postponing the progress of under-represented groups, while it eventually was accepted but such delays do impact their career. 

\begin{figure*}[t]
     \centering
     \includegraphics[width=\textwidth]{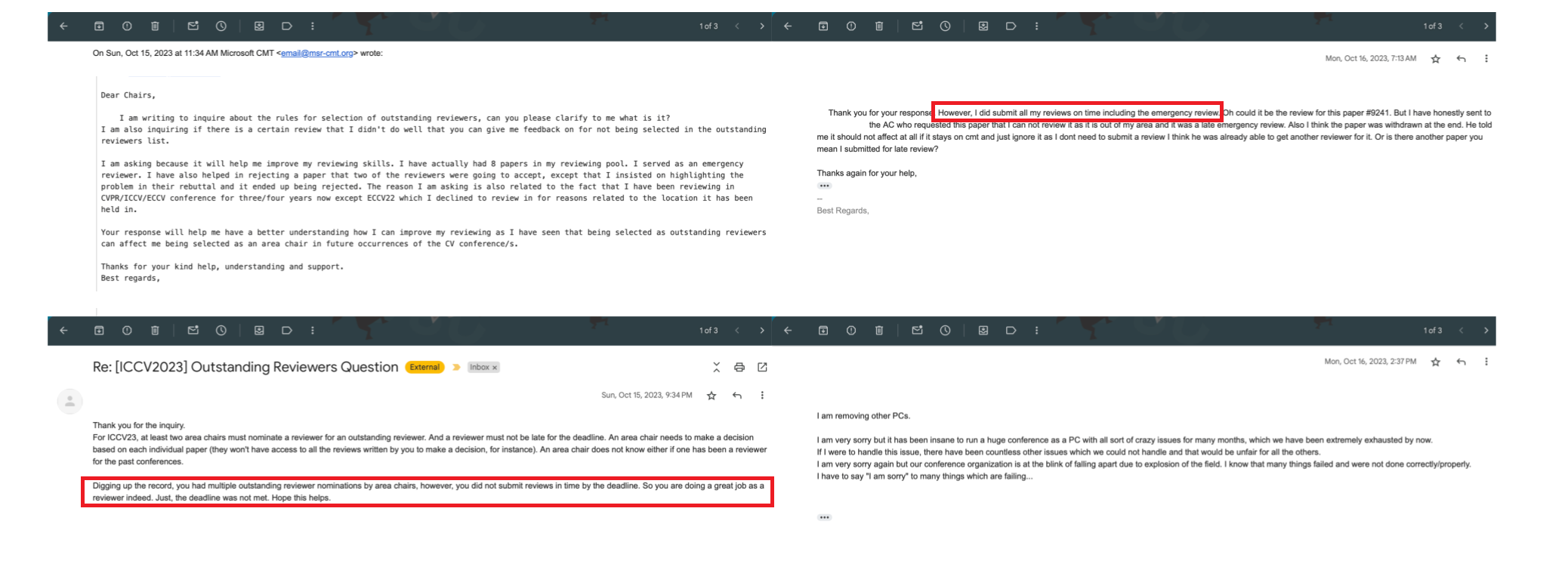}
     \vspace{-2em}
     \caption{The outstanding reviewer dispute, where the devil was denied the acknowledgment for a mistake.}
     \label{fig:oustanding}
     \vspace{-1em}
\end{figure*}

Another work~\footnote{\url{https://arxiv.org/abs/2307.07812}} that had a strong contribution and experiments, as it was evaluated across four tasks and on six datasets showing state of the art performance. It got rejected in Neurips 2022, CVPR 2023 and ICCV 2023. In one of the reviews she received, the reviewer claimed the baselines used were weak, Fig~\ref{fig:cvpr23}\footnote{\url{https://www.dropbox.com/scl/fo/rbrlt5kv4l09ollkg8q6c/ALjUwisjOhsP7X3hmvmRDuU?rlkey=29wmrh9zww0uv2gxqafw4cojh&st=csvxq0c1&dl=0}}. Except the baseline was outperforming the state of the art in this task from a work published in CVPR~\footnote{\url{https://openaccess.thecvf.com/content/CVPR2021/papers/Chen_Delving_Deep_Into_Many-to-Many_Attention_for_Few-Shot_Video_Object_Segmentation_CVPR_2021_paper.pdf}} and was re-implementing the stacked version from high quality recent research~\footnote{\url{https://arxiv.org/pdf/2112.01527}}. It is ok to reject the work but provide reasonable reviews that researchers can use to improve, not dead end type of reviews. Yet, another work~\footnote{\url{https://arxiv.org/pdf/2203.14308}} was rejected from CVPR 2022 and TPAMI, even when their proposed benchmark and developed algorithm was used in an ICCV 2023 work~\footnote{\url{https://arxiv.org/pdf/2309.11160}}. 

Some reviewers think they can kill papers because they have a crystal orb that can predict whether the authors of a paper are cheating or not, then use false claims to get rid of these works but with a good intention. False claims include, missing related work even when they exist, ambiguous concerns regarding writing, or others. Except, was it good intention though! Or is it that some reviewers want to ensure certain researchers are the ones accepted and others are rejected to reduce competition! According to the reviewing guidelines\footnote{\url{https://cvpr.thecvf.com/Conferences/2024/ReviewerGuidelines}}, the work is evaluated based on what is provided in the work and its supplementary material. Hence, trying to predict something without sufficient evidence, and even worse not providing this reason to the authors should never be an excuse to reject any work.

Finally, her recent work with her visiting student~\footnote{\url{https://arxiv.org/abs/2402.12519}} got rejected in ICLR 2023 and NeurIPS 2024. It is normal for works to be rejected initially and can help the work to improve, except the NeurIPS rejection killed the whole research direction with an unfounded claim, Fig.~\ref{fig:neurips2024}\footnote{\url{https://www.dropbox.com/scl/fo/4a94h7ui2irsygmwa8otw/AKVLTqOXMSXcXnmHuyZGB3o?rlkey=d9bcr9aocncf5t0fpdyhk676c&st=fi7ueblw&dl=0}}. It stated the dampened enthusiasm in studying video stimuli with fMRI, even when there are recent works that were published in Nature Communications 2024~\footnote{\url{https://www.nature.com/articles/s41467-024-50310-3}}. The student herself was discouraged if the review is killing the whole direction whether to continue in this direction or not. Imagine an African female student that is affiliated with an African institution working day and night on work that showed sufficient novelty, large-scale experimental results, showing error bar plots and statistical significance, showing both simulated and real experiments. Yet, the reviewer kills the whole direction not even leaving room to improve the work. Is this sufficient evidence that the conference reviewing system is broken? Not yet! 

Aside from the devil as an author, the devil as a reviewer was facing issues there too which is quite alarming. The devil sent to the program chairs in ICCV 2023 to inquire how to improve her reviewing and how they identify outstanding reviewers. The reason is that she served as an emergency reviewer and successfully promoted good research while rejecting low quality research that other reviewers were about to accept. Indeed she received an email from one of the program chairs saying she got multiple nominations from ACs for being an outstanding reviewer but it ended up being a mistake for one review being identified as late review when it was not, Fig.~\ref{fig:oustanding}. She requested for the mistake to be fixed and that she is added to the list of outstanding reviewers, yet that request was denied. She had to go to the Ombuds to fix the problem. The question that emerges if there is a fight for every single acknowledgment even the minor ones for minorities, is this sufficient to claim we have a broken system. Is our reviewing system truly driven by evaluating the quality of the work itself solely, or do grants, collaborations and reputation affect it? Some argue that reputation is important, except if it is the drive behind papers getting accepted it is cheating. 

This is not to undermine the efforts of the organizing committees, area chairs and reviewers and there have been positive improvements including: (i) CVPR 2024 allowing an appeal process, while all were rejected, it was still a good process since it allowed for another phase to identify extreme cases of bad reviews which did not occur. (ii) CVPR 2025 allowing reviewer names to be visible to other reviewers at the end of the process to discourage sloppy reviews. (iii) ICLR 2025 including a chat agent to identify and recommend changes to the reviews when fit. Even better, if we can create chat agents recommending to reviewers their consistency in their reviews among papers to prevent cases of favoring or discrimination. More importantly, we suggest the creation of an ethics committee to monitor the reviewing process with the first duty to collect evidence of false claims in reviews from the past 2-3 years from all researchers to study how our system is broken. Let's conduct proper review of the issue before solving the problem.

\section{Spreading Lies}
\label{sec:rumours}
\epigraph{\textbf{TL;DR. We used to be scientists working with evidence until we decided to be popular.}}


In this section, we discuss some of the reasons that might have contributed to our researcher being labeled as the devil beyond the reprisal and discrimination. She faced the label of; ``She is a terrorist'' which is a typical stereotype pointed to a Muslim woman. Except if she was she would be in jail right now, so how this label is still persistent is unknown. Another one, ``She is working with military and police in her home country'' that would be the impossible since she has been shot at from these and witnessed a massacre herself. There is also no reason for someone that educated, with a good job and settled in another country to do that. 

Other claims on the scientific side include; ``She is cheating and fabricating results'', when all her papers have the code released and reproducible. So if there are issues, why the researchers did not simply send to the devil herself or open an issue regarding the problem. Another one, ``She is stealing ideas''. Quoting, Yejin Choi from her talk in CVPR 2023, good ideas stem from researchers who are brave enough to come up with it and are not afraid. Their only constraint is scientific rules and standards not how people will perceive them. The people spreading lies behind others' back are not the ones who are brave. Generally speaking, any allegations that are being spread should simply be conveyed to the ones suspected to have committed these allegations and given the opportunity to defend themselves. If they are found to have committed harm to the community then that make sense to spread the truth in that case with full transparency.

Along that line, the devil's research ideas included the following; She was working on exploring the fusion of foundation models and test-time adaptation ones, the research direction was to account for situations where generalist approaches alone would fall short. Take an example of video anomaly detection, a generalist approach would be capable of certain aspects of it. But what is even considered an anomaly, changes from one scene, time and location to another. She also wanted to explore the ability of multi-modal large language models to be used for assessing datasets in self-supervised or vision language modeling settings beyond the need of conducting the full training with each change in the dataset. What certain dataset attributes that do correlate with the models' ability to generalize. This is in pursuit of a method that does not have a detrimental impact on Climate change with each company or research team training their own foundation model/s. Furthermore, she wanted to study the views on Muslim women especially Hijabi women in short videos and how these can correlate to widely spread stereotypical views. She was also studying how pixel-level foundation models such as segment anything define an object, she was studying that from the lens of Gestalten principles. Except the devil was prevented from hiring students to work on her research! 

``She is using Africans'' or ``She is abusing her power and killing young researchers''. Except the devil has never been funded for flights or hotel stays from any of the diversity communities including Deep Learning Indaba or Black in AI based on her request. Since, she preferred others to benefit from this. The only time she was funded as a PhD student, was back in 2017 NeurIPS from WiML. In fact, she shared her hotel room with two black women organizers in the Black in AI Social in CVPR 2023. At the end, the powerful is not the one that is about to end up unemployed so the claim that this devil is powerful is unfounded. 


If spreading lies without evidence is sufficient these days, so let's spread this one. Yann Lecun, Geoffrey Hinton and Yohsua Bengio are mind controlling everyone on earth and conducting unethical human experiments that started back in Montreal~\cite{klein2007shock}. Goes without saying we are being cynical and that previous claim is not true, but let's try to be popular.

\section{Conclusions}
In conclusion, this work has documented the human rights violations that are ongoing within our community that are undermined by the majority of researchers. A relatively successful researcher with respect to her diversity group was prevented from starting her research lab. She was vilified without being provided evidence of any allegation or reason for this, and yes even a single researcher rights being violated still needs to be fixed. More importantly, large-scale human right violations like Genocides have to be condemned and is the first to be addressed to ensure inclusivity not just as a pretense. At the end, what we normalize now happening to others is what will happen to us in the future, so out of wisdom we are better off solving them now.

\small
\bibliographystyle{ieeenat_fullname}
\bibliography{main}

\end{document}


\makeatletter
\DeclareRobustCommand\onedot{\futurelet\@let@token\@onedot}
\def\@onedot{\ifx\@let@token.\else.\null\fi\xspace}
\def\eg{\emph{e.g}\onedot} \def\Eg{\emph{E.g}\onedot}
\def\ie{\emph{i.e}\onedot} \def\Ie{\emph{I.e}\onedot}
\def\cf{\emph{cf}\onedot} \def\Cf{\emph{C.f}\onedot}
\def\etc{\emph{etc}\onedot} \def\vs{\emph{vs}\onedot}
\def\wrt{w.r.t\onedot} \def\dof{d.o.f\onedot}
\def\etal{\emph{et al}\onedot}

\def\latentregion{latent region}
\def\latentregionrep{latent token}

\makeatother
\maketitle

\input{sec/X_suppl}

{
    \small
    \bibliographystyle{ieeenat_fullname}
    \bibliography{main}
}